\documentclass[11pt,letterpaper]{article}
\usepackage{amsmath,amssymb}

\author{W.~M\"uck\thanks{E-mail: \texttt{wmueck@sfu.ca}}~\
and K.~S.~Viswanathan\thanks{E-mail: \texttt{kviswana@sfu.ca}}\\
\emph{\small Department of Physics, Simon Fraser University, Burnaby, B.C.,
V5A 1S6 Canada}}
\title{Regular and Irregular Boundary Conditions\\ in the AdS/CFT
Correspondence}

\DeclareMathAlphabet{\mathbd}{OT1}{cmr}{b}{n}
\begin{document}

\providecommand{\e}{\mathrm{e}}
\providecommand{\bd}{\mathbd{d}\mspace{-1mu}}
\providecommand{\bvec}[1]{\ensuremath{\mathbd{#1}}}
\providecommand{\kint}{\int \frac{d^d\mspace{-2mu}k}{(2\pi)^d}\,}
\providecommand{\Ord}{\mathcal{O}}
\providecommand{\eikx}{\e^{-i\bvec{k\cdot x}}}
\providecommand{\dd}[1]{d^d\mspace{-2mu}{#1}\,}
\providecommand{\xlim}{\overset{x_0\to0}{\approx}}

\maketitle
\begin{abstract}
We expand on Klebanov and Witten's recent proposal for
formulating the AdS/CFT correspondence using irregular boundary
conditions. The proposal is shown to be correct to any order in
perturbation theory.
\end{abstract}
\newpage

\section{Introduction}
\label{intro}

The celebrated AdS/CFT correspondence relates field theories on
anti-de Sitter (AdS) space with conformal field theories (CFTs) living
on the AdS horizon. The main prediction of this duality is that CFT
correlation functions of conformal operators can be calculated by
evaluating the AdS action on-shell as a functional of prescribed
boundary values. 

For example, using a scalar field theory on AdS space, CFT correlators
of conformal fields of scaling dimensions $\Delta\ge d/2$ have been
calculated \cite[and references therein]{Gubser98-1,Witten98-1,Mueck98-1,Freedman99-1,Liu99-1,dHoker99-2}.
Until recently, no prescription was known to include operators with scaling
dimension $\Delta$, $d/2-1<\Delta<d/2$. Here, $d/2-1$ is the unitary bound on
the conformal dimension of scalar operators. 
Recently, Klebanov and Witten \cite{Klebanov99-1} proposed a method to
do just that. They used the fact that a scalar field on AdS space can
obey two types of boundary conditions \cite{Breitenlohner82}. 
The regular one, which can always be imposed, leads to the CFT
correlators with $\Delta\ge d/2$, whereas the irregular one would lead
to $d/2-1<\Delta\le d/2$. They realized that the respective boundary
fields are conjugate to each other and proposed to use a Legendre
transform of the action, expressed as a functional of the irregular
boundary value, as the generating functional. They also demonstrated
the correctness of this proposal for CFT two point functions.

In this article, we would like to expand on their proposal and
demonstrate its correctness to all orders in perturbation theory. A
second result of our analysis is that a different Green's function must
be used for internal lines in second or higher order graphs. 

The outline of the article shall be as follows. In the remainder of
this section motivating arguments about the origin of the irregular
boundary conditions will be given. In section~\ref{reg} we will for
completeness repeat the formalism using regular boundary
conditions. Then, in section~\ref{irreg} Klebanov and Witten's
proposal to include irregular boundary conditions shall be analyzed
and shown to be correct to any order in perturbation theory.

To start, consider an interacting scalar field, whose action is given by 
\begin{equation}
\label{intro:action}
  I = \frac12 \int_\Omega \bd x\, \left(D\mspace{-2mu}_\mu \phi D^\mu
  \phi + m^2 \phi^2 \right) + I_{int},
\end{equation}
where $I_{int}$ denotes the interaction terms and $\bd x=
d^{d+1}\!x\sqrt{g(x)}$ is the invariant volume integral measure. The
equation of motion following from the action \eqref{intro:action} is
given by 
\begin{equation}
\label{intro:eqmot}
  \left( D\mspace{-2mu}_\mu D^\mu -m^2 \right) \phi (x) = B(x),
\end{equation}
where\footnote{The functional variation is done covariantly, cf.\
\cite{Basler93}.}  
\[ B(x) = \frac{\delta I_{int}}{\delta \phi(x)}.\]
Using as AdS representation the conventional upper half space
$\bvec{x}\in\mathbb{R}^d$, $x_0>0$ with the metric 
\begin{equation}
\label{intro:metric}
  ds^2 = (x_0)^{-2} dx^\mu dx^\mu,
\end{equation}
the solution to equation \eqref{intro:eqmot} can be written in the
form
\begin{equation}
\label{intro:sol}
  \phi(x) = \int \dd{y} \left[ \frac{x_0}{(x-\bvec{y})^2}
  \right]^{\frac{d}2\pm\alpha} f_\mp(\bvec{y}) + 
  \int_\Omega \bd y\, G(x,y) B(y),
\end{equation}
where $\alpha=\sqrt{d^2/4+m^2}$ and $G(x,y)$ is a standard Green's
function satisfying 
\begin{equation}
\label{intro:G}
  \left( D\mspace{-2mu}_\mu D^\mu -m^2 \right) G(x,y) =
  \frac{\delta(x-y)}{\sqrt{g(x)}}. 
\end{equation}
The free field solution with the lower sign exists classically for
$\alpha<d/2$, but the unitary bound restricts it further to $\alpha<1$
\cite{Breitenlohner82}. The functions $f_-$ and $f_+$ are
called regular and irregular boundary values and are
conformal fields of scaling dimensions $d/2-\alpha$ and $d/2+\alpha$,
respectively. 

In the AdS/CFT correspondence the fields obeying regular boundary
conditions give rise to CFT correlation functions of operators with
conformal dimensions $\Delta$ restricted by $\Delta\ge d/2$. Hence,
the use of irregular boundary conditions enables one to obtain
correlation functions for operators with scaling dimensions
$d/2-1<\Delta<d/2$. 

\section{Regular Boundary Conditions}
\label{reg}
Let us start by rewriting the expression \eqref{intro:sol} as 
\begin{equation}
\label{reg:phi} 
  \phi(x) = \phi^{(0)}(x) + \int_\Omega \bd x\, G(x,y) B(y),
\end{equation}
where the Green's function $G(x)$ is given by
\cite{Burgess85,Mueck98-1}
\begin{align}
\label{reg:Gans}
  G(x,y) &= -(x_0y_0)^\frac{d}2 \kint \e^{-i\bvec{k\cdot(x-y)}}
  \begin{cases} I_\alpha(kx_0) K_\alpha(ky_0) &\text{for $x_0<y_0$},\\
                I_\alpha(ky_0) K_\alpha(kx_0) &\text{for $x_0>y_0$},
  \end{cases} \\
\label{reg:G}
  &= - \frac{c_\alpha}2 \xi^{-\left(\frac{d}2+\alpha\right)}
  F\left(d/2,d/2+\alpha;1+\alpha;\xi^{-2}\right),
\end{align}
where $F$ is the hypergeometric function, 
\[ \xi = \frac1{2x_0y_0} \left\{ \frac12
 \left[ (x-y)^2 +(x-y^*)^2 \right] +\sqrt{(x-y)^2(x-y^*)^2} \right\} \]
($y^*$ denotes the vector $(-y_0,\bvec{y})$), and 
\begin{equation}
\label{reg:calpha}
  c_\alpha = \frac{\Gamma(d/2+\alpha)}{\pi^{d/2}\Gamma(1+\alpha)}.
\end{equation}

Moreover, the free field solution $\phi^{(0)}$ shall be written as
\begin{equation}
\label{reg:phi0}
  \phi^{(0)}(x) 
  = \int\dd{y}\mathcal{K}_\alpha(x,\bvec{y}) \phi_-^{(0)}(\bvec{y})
  = \int\dd{y}\mathcal{K}_{-\alpha}(x,\bvec{y})\phi_+^{(0)}(\bvec{y}).
\end{equation}
The bulk-boundary propagators occuring in equation \eqref{reg:phi0} 
are given by
\begin{equation}
\label{reg:K}
  \mathcal{K}_{\pm\alpha}(x,\bvec{y}) = \pm\alpha c_{\pm\alpha} \left[
  \frac{x_0}{(x-\bvec{y})^2}\right]^{\frac{d}2\pm\alpha},
\end{equation}
where $c_{\pm\alpha}$ is given by equation \eqref{reg:calpha}, and their
Fourier transforms read
\begin{equation}
\label{reg:Kk}
  \mathcal{K}_{\pm\alpha}(x,\bvec{k}) =
  \frac{\pm2\alpha}{\Gamma(1\pm\alpha)} \e^{i\bvec{k\cdot x}}
  \left(\frac{k}2\right)^{\pm\alpha} x_0^\frac{d}2 K_\alpha(kx_0).
\end{equation} 
Equations \eqref{reg:Kk} and \eqref{reg:phi0} imply that the
boundary functions $\phi^{(0)}_+$ and $\phi^{(0)}_-$ are related by
\begin{equation}
\label{reg:phi0pm}
  \phi^{(0)}_+(\bvec{k}) = -\frac{\Gamma(1-\alpha)}{\Gamma(1+\alpha)} 
  \left(\frac{k}2\right)^{2\alpha} \phi^{(0)}_-(\bvec{k}).
\end{equation}

Obviously, the free field $\phi^{(0)}$ can be
written as a sum of two series, whose leading powers are
$x_0^{d/2-\alpha}$ and $x_0^{d/2+\alpha}$, respectively. Thus, one
finds by direct comparison with equations \eqref{reg:phi0} and
\eqref{reg:Kk} that the small $x_0$ behaviour of
$\phi^{(0)}$ is  
\begin{equation}
\label{reg:phi0x0}
  \phi^{(0)}(x) \xlim x_0^{\frac{d}2-\alpha} \phi_-^{(0)}(\bvec{x}) + 
  x_0^{\frac{d}2+\alpha} \phi_+^{(0)}(\bvec{x}),
\end{equation}
where subleading terms have been dropped.
Moreover, the Green's function \eqref{reg:G} goes like 
\begin{equation}
\label{reg:gx0}
  G(x,y) \xlim -\frac1{2\alpha} x_0^{\frac{d}2+\alpha}
  \mathcal{K}_\alpha(\bvec{x},y).
\end{equation}
Hence, the interaction contributes only to the $\phi_+$ part of the
asymptotic boundary behaviour, i.e.\ one can write 
\begin{align}
\label{reg:phix0}
  \phi(x) &\xlim x_0^{\frac{d}2-\alpha} \phi_-(\bvec{x}) +
  x_0^{\frac{d}2+\alpha} \phi_+(\bvec{x}),\\
\intertext{where}
\label{reg:phim}
  \phi_-(\bvec{x}) &= \phi_-^{(0)}(\bvec{x}),\\
\label{reg:phip}
  \phi_+(\bvec{x}) &= \phi_+^{(0)}(\bvec{x}) - \frac1{2\alpha}
  \int_\Omega \bd y\, \mathcal{K}_\alpha(\bvec{x},y) B(y).
\end{align}
Identical relations hold for the Fourier transformed expressions. 

Now consider the on-shell action, treated as a functional of the
regular boundary values $\phi_-$. Integrating equation
\eqref{intro:action} by parts yields
\[I = \frac12 \int \dd{x} x_0^{-d} n^\mu \phi \partial_\mu \phi -
  \frac12 \int_\Omega \bd x\, \phi(x) B(x) + I_{int}.\]
The first term must be regularized, which is done by writing
\begin{align*} 
 x_0^{-d} n^\mu \phi \partial_\mu \phi &= - x_0^{-d} \phi \left[
 \left(\frac{d}2-\alpha\right) x_0^{\frac{d}2-\alpha} \phi_- +
 \left(\frac{d}2+\alpha\right) x_0^{\frac{d}2+\alpha} \phi_+
 +\cdots\right]\\
 &= -x_0^{-d} \left(\frac{d}2-\alpha\right) \phi^2 - 2 \alpha \phi_-
 \phi_+ +\cdots,
\end{align*}
where the ellipses indicate contributions from subleading
terms and other terms which vanish for $x_0=0$. 
The first term in the last line is cancelled by a covariant
counterterm. Hence, the renormalized on-shell action is
\begin{align}
\notag
  I[\phi_-] &= -\alpha \kint \phi_-(\bvec{k}) \phi_+(\bvec{-k}) -
  \frac12 \int_\Omega \bd x\, \phi(x)B(x) +I_{int}\\
\label{reg:I}
  &= I^{(0)}[\phi_-]
  - \frac12 \int_\Omega \bd x\, \bd y\; B(x) G(x,y) B(y) + I_{int},
\end{align}
where equations \eqref{reg:phip}, \eqref{reg:phim},
\eqref{reg:phi0pm}, \eqref{reg:phi} and \eqref{reg:phi0} have
been used. The term $I^{(0)}$ in equation \eqref{reg:I} is given by 
\begin{align}
\notag
  I^{(0)}[\phi_-] &= \alpha \frac{\Gamma(1-\alpha)}{\Gamma(1+\alpha)}
  \kint\left(\frac{k}2\right)^{2\alpha} \phi_-(\bvec{k})
  \phi_-(\bvec{-k})\\ 
\label{reg:I0}
  &= - \alpha^2 c_\alpha \int \dd{x}\dd{y} \frac{\phi_-(\bvec{x})
  \phi_-(\bvec{y})}{|\bvec{x-y}|^{d+2\alpha}} 
\end{align}
and thus yields the correct two point function of scalar operators of
conformal dimension $\Delta = d/2+\alpha$, if one uses the AdS/CFT
correspondence formula
\begin{equation}
\label{reg:corr}
  \e^{-I[\phi_-]} = \left\langle \exp\left[ \alpha \int \dd{x}
  \mathcal{O}(\bvec{x}) \phi_-(\bvec{x}) \right] \right\rangle.
\end{equation}
The other two terms have to be expressed as a perturbative
series in terms of $\phi^{(0)}$. However, by virtue of equations
\eqref{reg:phi0} and \eqref{reg:phim} this naturally yields a
perturbative series in terms of the boundary function $\phi_-$.

\section{Irregular Boundary Conditions}
\label{irreg}
The treatment of irregular boundary conditions follows an idea by
Klebanov and Witten \cite{Klebanov99-1}. Consider the expression
\begin{align*}
 \frac{\delta I[\phi_-]}{\delta\phi_-(\bvec{k})} &= 2 \alpha
 \frac{\Gamma(1-\alpha)}{\Gamma(1+\alpha)}
 \left(\frac{k}2\right)^{2\alpha} \phi_-(-\bvec{k}) + \int_\Omega \bd
 x\, B(x) \frac{\delta\phi(x)}{\delta\phi_-(\bvec{k})}\\
 &\quad - \int_\Omega \bd x\, \bd y\, \bd z\, \frac{\delta^2
 I_{int}}{\delta\phi(x)\delta\phi(z)} G(x,y) B(y)
 \frac{\delta\phi(z)}{\delta\phi_-(\bvec{k})}.\\
\intertext{Using equation \eqref{reg:phi0pm} and the formula}
 \frac{\delta\phi(x)}{\delta\phi_-(\bvec{k})} &=
 \mathcal{K}_\alpha(x,-\bvec{k}) + \int_\Omega \bd y\,\bd z\, G(x,y)  
 \frac{\delta^2 I_{int}}{\delta\phi(y)\delta\phi(z)} 
 \frac{\delta\phi(z)}{\delta\phi_-(\bvec{k})},
\end{align*}
one finds
\[ \frac{\delta I[\phi_-]}{\delta\phi_-(\bvec{k})} = -2 \alpha
 \phi_+^{(0)}(\bvec{-k}) + \int_\Omega \bd x\,
 \mathcal{K}_\alpha(x,\bvec{-k}) B(x) = -2\alpha \phi_+(\bvec{-k}), \]
or, after an inverse Fourier transformation,
\begin{equation}
\label{irreg:delIphi}  
  \frac{\delta I[\phi_-]}{\delta\phi_-(\bvec{x})} = -2\alpha
  \phi_+(\bvec{x})
\end{equation}
This expression holds to any order in perturbation theory. This fact was
obtained in \cite{Klebanov99-1} using graph arguments. 
Furthermore, it shows first that $\phi_+$ can be regarded as the
conjugate field of $\phi_-$ and secondly that the functional 
\begin{equation}
\label{irreg:Jdef}
  J[\phi_-,\phi_+] = I[\phi_-] + 2\alpha \int \dd{x} \phi_-(\bvec{x})
  \phi_+(\bvec{x}) 
\end{equation}
has a minimum with respect to a variation of $\phi_-$. 

Klebanov and
Witten's idea \cite{Klebanov99-1} is to formulate the AdS/CFT
correspondence by the formula 
\begin{equation}
\label{irreg:corr}
  \e^{-J[\phi_+]} = \left\langle \exp\left[ \alpha \int \dd{x}
  \mathcal{O}(\bvec{x}) \phi_+(\bvec{x}) \right] \right\rangle.
\end{equation}
Here, the functional $J[\phi_+]$ is a Legendre transform of the action $I$,
i.e.\ it is the minimum value of the expression \eqref{irreg:Jdef},
expressed in terms of $\phi_+$.  

In the following, Klebanov and Witten's result about the correctness of
the two point function \cite{Klebanov99-1} shall be confirmed and
interactions included. The minimum of $J$ is easiest found from equations
\eqref{reg:I} and \eqref{irreg:Jdef}, giving
\begin{align}
\notag
  J[\phi_+] &= \alpha \kint \phi_-(\bvec{k}) \phi_+(\bvec{-k}) -
  \frac12 \int_\Omega \bd x\, \phi(x)B(x) +I_{int}\\
\notag
  &= - \alpha \frac{\Gamma(1+\alpha)}{\Gamma(1-\alpha)} \kint
  \left(\frac{k}2\right)^{-2\alpha} \phi_+(\bvec{k}) \phi_+(\bvec{-k}) 
  - \frac12 \int_\Omega \bd x\, \phi(x)B(x) +I_{int}\\  
\notag
  &\quad +\frac12 \int \dd{x} \int_\Omega \bd y\,
  \mathcal{K}_{-\alpha}(y,\bvec{x}) \phi_+(\bvec{x}) B(y)\\
\notag
  &= - \alpha \frac{\Gamma(1+\alpha)}{\Gamma(1-\alpha)} \kint
  \left(\frac{k}2\right)^{-2\alpha} \phi_+(\bvec{k}) \phi_+(\bvec{-k})
  -\frac12 \int_\Omega \bd x\, \bd y\; B(x) G(x,y) B(y) \\
\label{irreg:J1}
  &\quad + I_{int} -\frac1{4\alpha} \int \dd{z} \int_\Omega \bd x\,
  \bd y\; \mathcal{K}_\alpha(x,\bvec{z})
  \mathcal{K}_{-\alpha}(y,\bvec{z}) B(x) B(y).
\end{align}
Here equations \eqref{reg:phim}, \eqref{reg:phi0pm},
\eqref{reg:phip}, \eqref{reg:Kk} and \eqref{reg:phi} have been
used. The first term in equation \eqref{irreg:J1} can be inversely
Fourier transformed, which yields  
\begin{equation}
\label{irreg:J0}
  J^{(0)} = - \alpha^2 c_{-\alpha} \int \dd{x}\dd{y}
  \frac{\phi_+(\bvec{x})\phi_+(\bvec{y})}{|\bvec{x-y}|^{d-2\alpha}}. 
\end{equation}
According to the correspondence formula \eqref{irreg:corr}, this
yields the correct two point function of conformal operators
$\mathcal{O}$ of scaling dimension $\Delta=d/2-\alpha$.

Then, the second and fourth term in equation \eqref{irreg:J1} can be
combined by defining the Green's function
\begin{equation}
\label{irreg:gtildedef}
  \tilde G(x,y) = G(x,y) + \frac1{2\alpha} \int \dd{z}
  \mathcal{K}_\alpha(x,\bvec{z}) \mathcal{K}_{-\alpha}(y,\bvec{z}).
\end{equation}
This modified Green's function $\tilde G$ also satisfies equation
\eqref{intro:G}, because the second term in equation
\eqref{irreg:gtildedef} does not contribute to the
discontinuity. Moreover, using equations \eqref{reg:Gans} and
\eqref{reg:Kk} one finds
\begin{align*}
 \tilde G(x,y) &= - (x_0y_0)^\frac{d}2 \kint
 \e^{-i\bvec{k\cdot(x-y)}}\\
 &\quad \times \left[ \frac{2 K_\alpha(kx_0)
 K_\alpha(ky_0)}{\Gamma(\alpha)\Gamma(1-\alpha)} 
 + \begin{cases} 
 K_\alpha(ky_0) I_\alpha(kx_0) & \text{for $x_0<y_0$},\\
 K_\alpha(kx_0) I_\alpha(ky_0) & \text{for $x_0>y_0$}, \end{cases}
 \right] \\
 &= - (x_0y_0)^\frac{d}2 \kint \e^{-i\bvec{k\cdot(x-y)}} \begin{cases} 
 K_\alpha(ky_0) I_{-\alpha}(kx_0) & \text{for $x_0<y_0$},\\
 K_\alpha(kx_0) I_{-\alpha}(ky_0) & \text{for $x_0>y_0$}, \end{cases}
\end{align*}
which differs from equation \eqref{reg:Gans} only by interchanging
$\alpha$ and $-\alpha$. Hence, the result \eqref{reg:G} can be
taken over, yielding
\begin{equation}
\label{irreg:gtilde}
  \tilde G(x,y) = - \frac{c_{-\alpha}}2
  \xi^{-\left(\frac{d}2-\alpha\right)}\; F\left(
  d/2,d/2-\alpha;1-\alpha;\xi^{-2} \right).
\end{equation}

Thus, inserting equation \eqref{irreg:gtildedef} into equation
\eqref{irreg:J1} yields 
\begin{equation} 
\label{irreg:J}
  J[\phi_+] = J^{(0)}[\phi_+] -\frac12 \int_\Omega \bd x\,\bd y\;
  B(x) \tilde G(x,y) B(y) +I_{int}.
\end{equation}
Moreover, one can see from equation \eqref{irreg:gtilde} that for
small $x_0$ $\tilde G$ behaves as 
\begin{equation} 
\label{irreg:tgx0}
  \tilde G(x,y) \xlim \frac1{2\alpha} x_0^{\frac{d}2-\alpha}
  \mathcal{K}_{-\alpha}(\bvec{x},y).
\end{equation} 
Hence, writing 
\begin{equation}
\label{irreg:phi}
  \phi(x) = \int \dd{y} \mathcal{K}_{-\alpha}(x,\bvec{y})
  \phi_+(\bvec{y}) +\int_\Omega \bd y\, \tilde G(x,y) B(y), 
\end{equation}
the interaction contributes only to $\phi_-$. This in turn means that,
expressing $I_{int}$ and $B$ as a perturbative series and using
equation \eqref{irreg:phi}, the functional $J$ is naturally expressed
in terms of the irregular boundary value $\phi_+$. Moreover, it has
the expected form, in that it is obtained from equation
\eqref{reg:I} by replacing $\alpha$ with $-\alpha$ and $\phi_-$ with
$\phi_+$. An important point is that the Green's function $\tilde G$
must be used for the calculation of internal lines. 

Finally, by a calculation similar to that of the derivation of
equation \eqref{irreg:delIphi} one finds
\begin{equation}
\label{irreg:delJphi}
  \frac{\delta J[\phi_+]}{\delta \phi_+(\bvec{x})} = 2\alpha
  \phi_-(\bvec{x}).
\end{equation} 
This is a final confirmation of the fact that the fields $\phi_-$ and
$\phi_+$ are conjugate to each other. 

In conclusion, we have expanded on Klebanov and Witten's recent idea
for formulating the AdS/CFT correspondence using irregular boundary
conditions, showing it to give the expected answers to any order in
perturbation theory.

\section*{Acknowledgments}
This work was supported in part by a grant from NSERC. 
W.~M.\ is very grateful to Simon Fraser University for financial support.


\end{document}